\documentclass[%
 reprint,
 amsmath,amssymb,
 aps, prx,
 longbibliography,
 superscriptaddress,
 altaffilsymbol
]{revtex4-1}
\usepackage[left=2cm, right=2cm, top=3cm, bottom=3cm]{geometry}
\usepackage{amsmath}
\usepackage{amsopn}
\usepackage{amsfonts}
\usepackage{natbib,graphicx,url,times,layouts,xcolor}
\usepackage{bm}
\usepackage{txfonts} 
\usepackage{cases} 
\usepackage{placeins} 
\definecolor{blueviolet}{rgb}{0.2, 0.2, 0.6}

\newcommand{\bra}[1]{{\left\langle{#1}\right\vert}}
\newcommand{\ket}[1]{{\left\vert{#1}\right\rangle}}
\newcommand{\unit}{\ \mathrm}

\begin{document}
\title{Implementing a Universal Gate Set on a Logical Qubit Encoded in an Oscillator}
\author{Reinier W. Heeres}
\email{reinier@heeres.eu}
\altaffiliation{These authors contributed equally}
\author{Philip Reinhold}
\email{reinier@heeres.eu}
\altaffiliation{These authors contributed equally}
\author{Nissim Ofek}
\author{Luigi Frunzio}
\author{Liang Jiang}
\author{Michel H. Devoret}
\author{Robert J. Schoelkopf}
\affiliation{Departments of Physics and Applied Physics, Yale University, New Haven, Connecticut 06520, USA}

\maketitle
\textsf{\textbf{
A logical qubit is a two-dimensional subspace of a higher dimensional system,
chosen such that it is possible to detect and correct the occurrence of certain errors \cite{Terhal:2015ks}.
Manipulation of the encoded information generally requires arbitrary and precise control over
the entire system \cite{Zhang:2012hl,Nigg:2014eb}.
Whether based on multiple physical qubits or larger dimensional modes
such as oscillators, the individual elements in realistic devices will always
have residual interactions which must be accounted for when designing logical operations.
Here we demonstrate a holistic control strategy which exploits accurate
knowledge of the Hamiltonian to manipulate a coupled oscillator-transmon system.
We use this approach to realize high-fidelity (99\%, inferred), decoherence-limited operations
on a logical qubit encoded in a superconducting cavity resonator using four-component cat states
\cite{Mirrahimi:2014js,ofek2016demonstrating}.
Our results show the power of applying numerical techniques \cite{Khaneja:2005jm} to control
linear oscillators and pave the way for utilizing their
large Hilbert space as a resource in quantum information processing.
%
}}

Quantum error correction (QEC) aims at the creation of logical qubits whose
information storage and processing capabilities exceed those of its constituent parts.
Significant progress has been made toward quantum state preservation by
repeated error detection using stabilizer measurements
in trapped ions \cite{chiaverini2004realization,Nigg:2014eb},
nitrogen vacancy centers \cite{Cramer:2016gv},
and superconducting circuits \cite{Kelly:2015gi,riste2015detecting,corcoles2015demonstration}.
In order to go beyond storage and to manipulate the encoded information,
one must perform operations on the whole system in such a way
that it results in the desired transformation within the two-dimensional
subspace defining the logical qubit. Any encoding scheme will consist of multiple
interacting components where the system dynamics are not naturally confined within the logical subspace.
Therefore, implementing operations requires carefully tailored controls which
address each component of the system and manage their mutual interactions.
Recent efforts have achieved this level of control
and have demonstrated operations on
a 5 qubit code in nuclear spin ensembles \cite{Zhang:2012hl}
and a 7 qubit code in trapped ions \cite{Nigg:2014eb}.

An alternative to logical qubit implementations based on multiple two level systems
is to encode quantum information in
continuous variable systems or oscillators, for which there are several schemes \cite{Gottesman:2001jb,Michael:2016uu}.
In particular so called ``cat states'', which are superpositions of coherent
states, can be used as the logical states of an encoded qubit \cite{Mirrahimi:2014js}.
They are attractive because coherent states are eigenstates of the
photon annihilation operator ($\hat{a}$) and therefore single-photon loss induces simple,
tractable errors.

\begin{figure*}[tbph]
  \centerline{\includegraphics{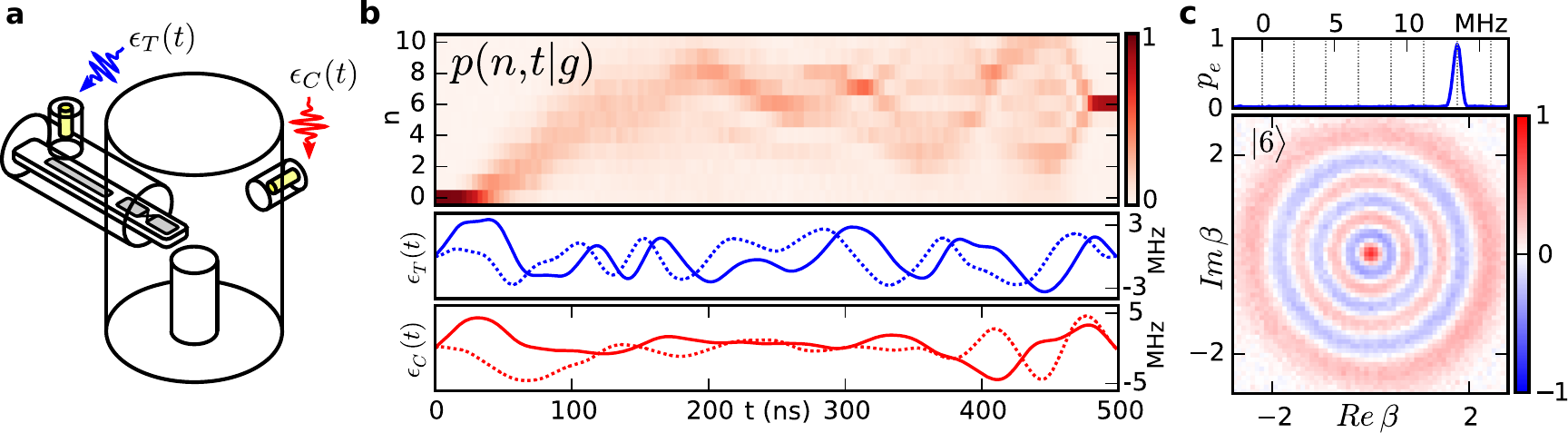}}
  \caption{\label{fig1}
  \textbf{Experimental system and demonstration of control strategy. a,}
  Schematic drawing of the experimental system. A $\lambda / 4$ coax-stub cavity resonator
  is coupled to a transmon and readout resonator on a sapphire substrate.
  Input couplers close to the transmon and cavity deliver the respective time-dependent
  microwave control fields $\epsilon_T(t)$ and $\epsilon_C(t)$.
  \textbf{b,} Lower panel: optimized transmon and oscillator control waveforms of length
  approximately $2\pi / \chi$ to take the oscillator from vacuum to the 6-photon Fock state.
  Solid (dotted) lines
  represent the in-phase (quadrature) field component. Upper panel: oscillator photon-number population
  trajectory versus time conditioned on transmon in $\ket{g}$. A complex trajectory occupying a wide range
  of photon numbers is taken to perform the intended operation. \textbf{c,} Characterization of the
  oscillator state using Wigner tomography (bottom) and transmon spectroscopy (top),
  where grey dashed lines indicate the transition frequency associated with
  the first 7 Fock states.
  The single peak in the spectroscopy data directly reveals the oscillator's
  population due to the dispersive interaction giving a frequency shift
  of $6\chi/2\pi \approx 13 \unit{MHz}$.
  }
\end{figure*}
Replacing several two level systems by an oscillator drastically reduces the hardware
cost and complexity  by requiring fewer components to fabricate and manipulate.
However, introducing higher dimensional modes raises the issue of how to realize
complete control over the system.
Driving an isolated harmonic oscillator results in a displacement operation,
which can only produce coherent states from the vacuum.
Any oscillator-based logical qubit scheme will require a richer class of operations,
which one can access via coupling to a nonlinear system.
In the case of a frequency-tunable qubit coupled to an oscillator with the Jaynes-Cummings (JC)
interaction ($H_\text{JC} = \sigma_+ \hat{a} + \sigma_- \hat{a}^\dagger$),
it has been demonstrated that it is possible to prepare arbitrary states in the
oscillator \cite{Law:1996cj,Hofheinz:2009ba}.

In the far off-resonant case, where the JC interaction reduces to the
dispersive Hamiltonian ($H_d/\hbar = \chi a^\dagger a \ket{e}\bra{e}$),
a small set of operations acting on a timescale of $2\pi/\chi$ is
in principle sufficient for universal control \cite{Krastanov:2015gn,Nigg:2014dl}
and has been used for non-trivial operations \cite{Vlastakis:2013ju,Heeres:2015kr}.
Generally, however, any approach decomposing an arbitrary operation
into a sequence of elementary gates generates only a small subset of physically
allowed control fields.
It therefore suffers from two issues limiting the achievable fidelity.
First, the constructed sequences may require an
unacceptably large number of gates, limiting operations which are
feasible in the presence of decoherence.
Second, the idealized model used by a constructive approach
typically fails to account for the existence of higher order Hamiltonian terms
such as the Kerr non-linearity
$H_\text{Kerr}/\hbar = \frac{K}{2}\left(\hat{a}^\dagger\right)^2 \hat{a}^2$
and spurious residual couplings in multi-qubit systems.


In this work we address these problems by considering
a full model of the time dependent Hamiltonian
in the presence of arbitrary control fields.
Nuclear magnetic resonance experiments have shown that,
if the available controls are universal, numerical optimization procedures can
reliably solve the inversion problem of finding
control fields to implement an intended operation.
These optimal control algorithms, in
particular the Gradient Ascent Pulse Engineering (\textsc{grape})
method \cite{Khaneja:2005jm,deFouquieres:2011wm},
have been successfully employed in a variety of other
fields \cite{dolde2014high,Anderson:2015fu}.
Since \textsc{grape} crucially depends on the model
of the system, its successful application is
powerful evidence that the Hamiltonian used accurately captures the
system dynamics over a broad range of driving conditions.

\begin{figure*}[tbph!]
  \centerline{\includegraphics{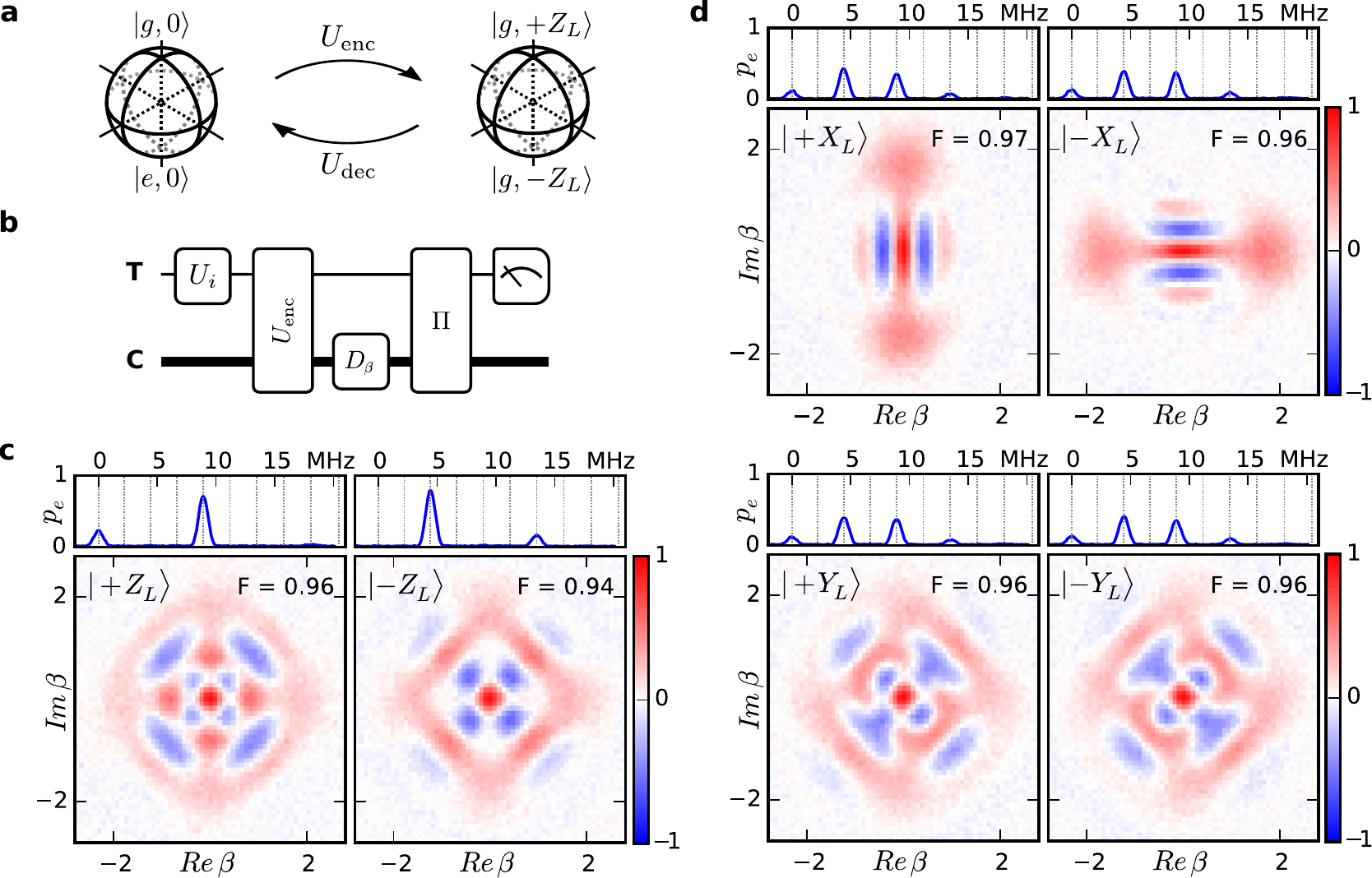}}
  \caption{\label{fig2} \textbf{Characterization of encoded states.}
  \textbf{a,} $U_\text{enc}$ and $U_\text{dec}$ are operations which coherently map
   between two distinct two-dimensional subspaces, represented by Bloch spheres. The first
   subspace consists of the transmon $\ket{g}$ and $\ket{e}$ levels, with the oscillator in the vacuum. The
   second is given by the oscillator-encoded states $\ket{+Z_L}$ and $\ket{-Z_L}$ (Eq. \ref{eq:codewords}),
   with the transmon in the ground state.
  \textbf{b,} Wigner tomography sequence which characterizes the encoded states. A transmon state is
   prepared by applying an initial rotation $U_i$ and is mapped to the oscillator using $U_\text{enc}$.
   An oscillator displacement $D_\beta$ followed by a parity mapping operation $\Pi$ (implemented
   using an optimal control pulse) allows one to measure the oscillator
   Wigner function $W(\beta)$.
   The transmon can be re-used to measure the oscillator's parity because the encoding
   pulse leaves the transmon in the ground state with high probability ($p > 98\%$).
  \textbf{c,} Applying $U_\text{enc}$ to the transmon states $\ket{g}$ and $\ket{e}$ produces
  states whose Wigner functions are consistent with the intended encoded basis states
  (Eq. \ref{eq:codewords}). A transmon spectroscopy experiment (top panel)
  illustrates that only photon number states with
  $n = 0\ \text{mod}\ 4$ and $n = 2\ \text{mod}\ 4$ are present for logical state
  $\ket{+Z_L}$ and $\ket{-Z_L}$ respectively.
  \textbf{d,} Applying $U_\text{enc}$ to superpositions of the transmon basis states
  demonstrates that the relative phase is preserved and that $U_\text{enc}$
  is a faithful map between the transmon and logical qubit Bloch spheres. These states,
  on the equator of the Bloch sphere, are equally weighted superpositions of
  $\ket{+Z_L}$ and $\ket{-Z_L}$ and therefore contain all even photon numbers present in
  the basis states.
}
\end{figure*}

The physical system used in our experiments is schematically depicted in
Fig.~1a. The seamless aluminum $\lambda / 4$ coax-stub cavity resonator \cite{MattCav}
with a fundamental frequency
$4452.6 \unit{MHz}$ has an energy relaxation time of $2.7 \unit{ms}$. A single-junction transmon
with transition frequency $5664.0 \unit{MHz}$
and anharmonicity of $236 \unit{MHz}$
is dispersively coupled to the oscillator, resulting in an interaction term
$\chi \hat{a}^\dagger \hat{a}\ket{e}\bra{e}$, with $\chi/2\pi = -2.2\unit{MHz}$.
Crucially, additional higher order terms are determined accurately through
a separate set of calibration experiments (Table SI, Supplementary Information).
Control of the system is
implemented through full in-phase/quadrature (IQ) modulated microwave fields
centered on the transmon (oscillator) frequency and sent to the pin coupling to the
transmon (oscillator) mode.  In the rotating wave approximation, this results in
the drive Hamiltonian $H_c/\hbar = \epsilon_C a + \epsilon_T \sigma_- + \text{h.c.}$
Modulation using an arbitrary waveform generator allows the coefficients
$\epsilon_C$ and $\epsilon_T$ to be arbitrary complex-valued functions of time.

As an example application of \textsc{grape} to our system,
we find $\epsilon_C(t)$ and $\epsilon_T(t)$ such that,
starting from the vacuum, after 500 ns of driven evolution
the system ends up in the state $\ket{g,6}$ (Fig.~1bc).
This highly nontrivial operation, effectively realizing a $\ket{6}\bra{0}$
coupling term on the oscillator, is enabled by the dispersive Hamiltonian
using only linear drives on the transmon and the oscillator.

\begin{figure}[tpbh]
  \mbox{\includegraphics{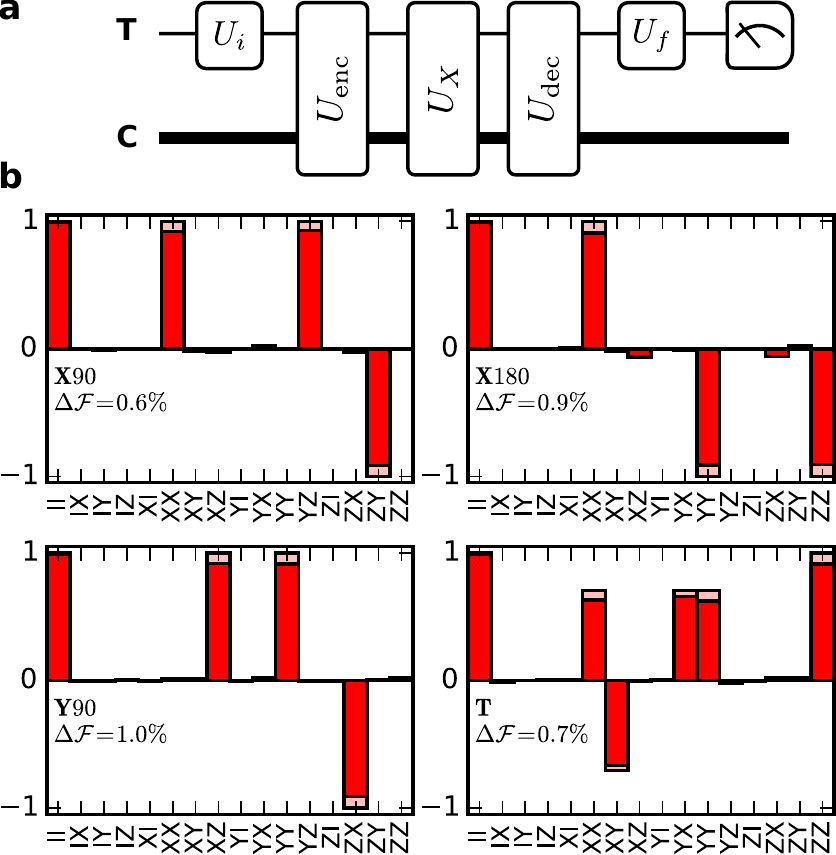}}
  \caption{\label{fig3}\textbf{Process tomography of operations on encoded qubit.}
  \textbf{a,} In order to characterize a gate $U_X$ on the encoded qubit, transmon process tomography
  is performed on the operation $U_\text{dec} U_X U_\text{enc}$. Process tomography
  is implemented by performing an initial transmon rotation $U_i$ right after state preparation,
  as well as a final transmon rotation $U_f$, right before measurement of the transmon.
  \textbf{b,} Process tomography results for selected operations (for additional operations,
  see Fig.~S7, Supplementary Information).
  The process tomography yields an estimated quantum channel $G$. We represent this channel
  in the Pauli transfer representation. The bar labeled with operators $AB$ represents
  $\text{Tr}\left(A G(B)\right) / 2$.
  Red and pink bars indicate the experimental and ideal values, respectively.
  The infidelity $\Delta \mathcal{F}_\text{PT}$ of operation $U_X$ is
  estimated as the difference between $\mathcal{F}_\text{PT}(U_\text{dec} U_X U_\text{enc})$
  and $\mathcal{F}_\text{PT}(U_\text{dec} U_\text{enc}) = 0.964$.
  The selected set of operations, $\left\{X180, X90, Y90, T \right\}$,
  allows universal control of the logical qubit.}
\end{figure}

Using this control strategy, we can target the creation and manipulation of a
logical qubit encoded in an even-parity four-component cat subspace.
Omitting normalization, the code states in this subspace can be written as
\begin{equation}
\label{eq:codewords}
\ket{\pm Z_L} = \ket{\alpha} + \ket{-\alpha} \pm \ket{i\alpha} \pm \ket{-i\alpha}
\end{equation}
where we use $\alpha = \sqrt{3}$. These code words are both
of even photon number parity, and are distinguished by their photon number
modulo 4:
\begin{align}
\ket{+Z_L} &= \sum_n \frac{\alpha^{4n}}{\sqrt{(4n)!}}\ket{4n}\\
\ket{-Z_L} &= \sum_n \frac{\alpha^{4n + 2}}{\sqrt{(4n+2)!}}\ket{4n+2}
\end{align}
Single photon loss, the dominant error channel
for the system, transforms both code words to states of odd photon number parity.
The encoded information, however, is preserved by this process as long
as one can keep track of the number of photons that have been lost.
Since parity measurements can be performed efficiently and non-destructively
\cite{sun2014tracking}, single photon loss is a correctable error \cite{ofek2016demonstrating}.

Using \textsc{grape}, we create a universal set of gates on our logical qubit,
which includes $X$ and $Y$ rotations by $\pi$ and $\pi/2$, as well as Hadamard and T gates.
These pulses are each $1100 \unit{ns} \approx 2.4\times 2\pi / \chi$ in length
with a 2 ns time resolution, although 99\% of the spectral content
lies within a bandwidth of 33 MHz (27 MHz) for the transmon (oscillator) drive
(Fig.~S2, Supplementary Information).
Each operation is designed to begin and end with the transmon in the ground state.
Additionally, we create encode ($U_\text{enc}$) and decode ($U_\text{dec}$) pulses
to transfer a bit of quantum information between our transmon
$\{\ket{g,0},\,\ket{e,0}\}$ subspace, which we can easily prepare
and measure, and our encoded subspace $\{\ket{g,+Z_L},\,\ket{g,-Z_L}\}$ (Fig.~2a).

We characterize the encode operation by preparing all 6 cardinal points
on the transmon Bloch sphere, applying the
encode pulse and performing Wigner tomography on the oscillator (Fig.~2b--d).
Maximum likelihood reconstruction of the density matrix associated with the measured
Wigner functions indicates an average fidelity of 0.96.
This metric underestimates the fidelity of $U_\text{enc}$
because it is affected by several sources of error not intrinsic
to the encoding operation itself, including error in the parity mapping and
measurement infidelity.

Process tomography provides a full characterization of
a quantum operation, but depends on pre-existing trusted operations
and measurements which are not available for our encoded subspace.
However, an indirect characterization of a gate $U_X$ on our
logical qubit can be performed using the operation $U_\text{dec} U_X U_\text{enc}$,
which maps the transmon subspace onto itself.
This allows one to use the trusted state preparations and measurements on the transmon
to perform tomography on the composite process (Fig.~3a).
The reconstructed process matrices show qualitative agreement
with the intended encoded qubit gates.
We can break the calculated infidelity down into 3 parts:
transmon preparation and measurement error, encode-decode error and gate error.
Using the experimentally determined process fidelities both without any operation
$\mathcal{F}_\text{PT}(\text{No Op.}) = 0.982$,
as well as with the encode and decode pulses $\mathcal{F}_\text{PT}(U_\text{dec}U_\text{enc}) = 0.964$,
we estimate an infidelity contribution
of approximately 1.8\% for each of the first two components.
To account for these factors to first order,
the infidelity of operations on the encoded qubit are
reported relative to $\mathcal{F}_\text{PT}(U_\text{dec}U_\text{enc})$.
We find an average infidelity of $0.75\%$ over our set of 9 gates (Table~I).

\begin{figure*}[tbph]
  \mbox{\includegraphics{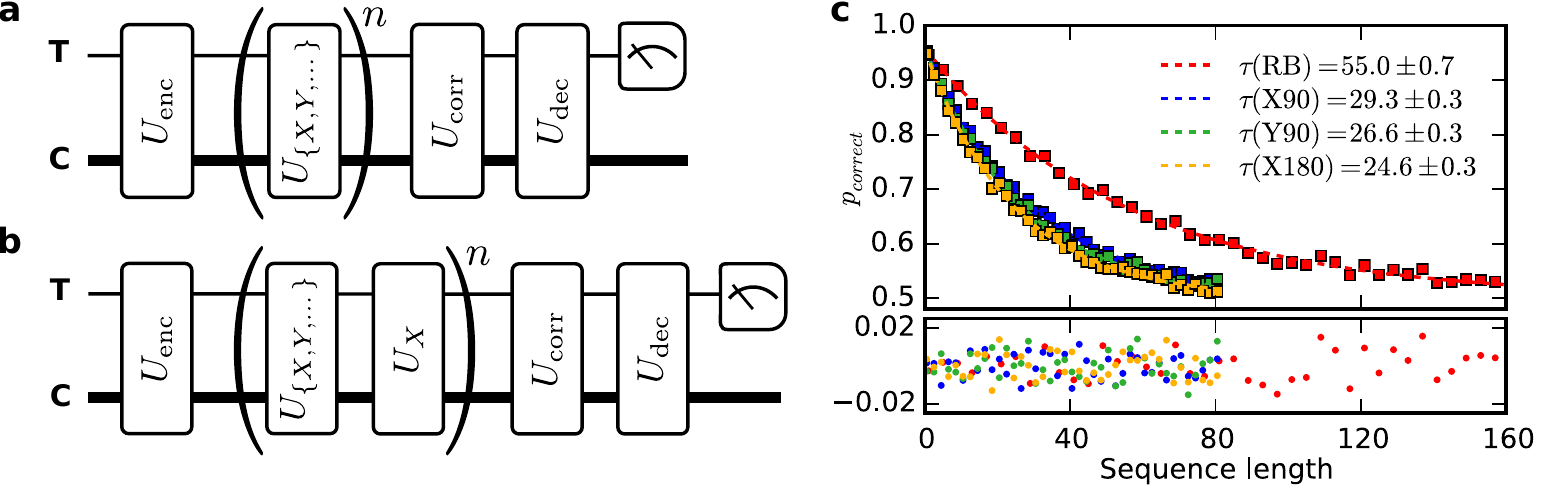}}
  \caption{\label{fig4}\textbf{Randomized benchmarking of operations on encoded qubit.}
  \textbf{a,} Randomized benchmarking (RB) sequence. In RB a sequence of
  Clifford operations of length $n$ is chosen at random ($U_{\{X,Y,\ldots\}}$),
  followed by the operation which inverts the effect of the sequence ($U_\text{corr}$).
  In order to apply this technique to the operations on the encoded qubit, we begin
  the experiment by encoding, and decode before measurement.
  Our implementation of RB creates a new random gate
  sequence for every measurement, and is thus not biased by the
  distribution of sequences which are measured.
  \textbf{b,} Interleaved randomized benchmarking (iRB) sequence:
  In order to establish the fidelity of a single operation (here, $U_X$),
  the operation is interleaved with random operations, and the benchmarking
  result is compared with the non-interleaved case.
  \textbf{c,} The probability of measuring the correct result
  versus sequence length $n$ is fit
  to a two parameter model $p_\text{correct} = 0.5 + A e^{-n/\tau}$.
  The lower panel shows the fit residuals.
  Each data point is the result of 2000 averages, with a new sequence realization every shot.
  The error averaged over all gates is computed as
  $r = (1 - e^{-1/\tau(\text{RB})})/2$ \cite{magesan2011scalable}.
  The average error for a single gate $X$ is computed as
  $r(X) = (1 - e^{1/\tau(X) - 1/\tau(\text{RB})})/2$ \cite{Magesan:2012gu}.
  }
\end{figure*}

In order to establish the fidelity of this set of operations more accurately, we perform
randomized benchmarking \cite{magesan2011scalable} (RB) on our encoded qubit (Fig.~4a).
From the resulting data (Fig.~4c) we infer an average gate fidelity of $0.991$.
The infidelity of each of the individual gates is isolated using
interleaved randomized benchmarking \cite{Magesan:2012gu} (iRB),
which alternates between a single fixed and a random gate (Fig.~4b). Comparing
the fitted decay constants of the RB and iRB results allows us to
extract the fidelity of the fixed gate. The results are summarized in
Table~\ref{fid_table}, together with the gate fidelities based on
process tomography (Fig.~3) and Lindblad master equation simulations
accounting for finite $T_1$  and $T_2$ of the transmon and oscillator
(see Supplementary Information).
We note that all gates are implemented with an approximately equal
infidelity of $1\%$ and that process tomography and iRB yield consistent results.
While several sources of decoherence are accounted for in the master equation
simulations, the dominant source of infidelity in the model is transmon
dephasing ($T_2 \approx 43 \unit{\mu s}$).
The strong agreement between simulations and experiment
indicates that the infidelity is primarily caused by decoherence
and that additional contributions associated with imperfections in the
model Hamiltonian and the applied pulses are a significantly smaller effect.

\begin{table}
\begin{tabular}{l c c c}
Gate   & $1 - \mathcal{F}_\text{RB}$ (\%) & $\Delta \mathcal{F}_\text{PT}$ (\%) & $1 - \mathcal{F}_\text{sim}$ (\%)\\
\hline
     I  & $0.46 \pm 0.02$ &  0.51   &   0.31 \\
   X90  & $0.79 \pm 0.02$ &  0.57   &   0.78 \\
  -X90  & $0.91 \pm 0.03$ &  0.71   &   0.83 \\
  X180  & $1.11 \pm 0.03$ &  0.88   &   1.09 \\
   Y90  & $0.96 \pm 0.03$ &  0.98   &   0.76 \\
  -Y90  & $0.81 \pm 0.02$ &  0.52   &   0.75 \\
  Y180  & $1.28 \pm 0.03$ &  0.99   &   1.67 \\
     H  & $0.93 \pm 0.03$ &  0.86   &   1.00 \\
	average & $0.90 \pm 0.02$ & 0.75 & 0.90\\

\hline
$U_\text{enc}U_\text{dec}$ & $1.70 \pm 0.03$ &  1.39   &  1.76 \\
     T & - &  0.71   &   0.40 \\
\end{tabular}
\caption{\label{fid_table} \textbf{Operation fidelities.}
Measured and simulated gate infidelities.
All fidelities reported are average gate fidelities
$\mathcal{F}(\mathcal{E}_1, \mathcal{E}_2) \equiv
\int\mathrm{d}\psi F(\mathcal{E}_1(\psi), \mathcal{E}_2(\psi))$,
where $F$ is the usual quantum state fidelity $F(\rho_1,\rho_2)=\text{Tr}(\sqrt{\rho_1} \rho_2 \sqrt{\rho_1})$.
$\mathcal{F}_\text{RB}$, $\Delta \mathcal{F}_\text{PT}$ and $\mathcal{F}_\text{sim}$
are the values extracted from interleaved randomized benchmarking,
process tomography (see Fig.~3) and simulations using the Lindblad
master equation respectively.
The row labeled ``average'' gives the fidelities averaged over the first
8 gates, which is the set used in the standard randomized benchmarking experiment.
}
\end{table}

In conclusion, we have demonstrated a high-fidelity implementation
of a universal set of gates on a qubit encoded into an oscillator using the cat-code.
The low error rates for these operations are verified using both process tomography
and randomized benchmarking, and the results are consistent with simulations
which account for decoherence.
We obtained these operations by numerically optimizing time-dependent drives
which make use of the well-characterized dispersive interaction between the
far detuned oscillator and transmon modes.
While in this Letter we have focused on realizing and
characterizing single-qubit operations on cat-encoded qubits,
this control technique is not restricted to these goals, and is
in principle capable of crafting arbitrary unitary operations
on the transmon-oscillator system.
The high quality of these operations depends critically on an accurate characterization of the
system Hamiltonian, and demonstrates the utility of numerical optimal control for realizing quantum information processing.


\section{Acknowledgments}
We would like to thank Katrina Sliwa and Michael Hatridge for providing the
parametric amplifier, Chris Axline, Jacob Blumoff, Kevin Chou and Chen Wang
for discussions regarding sample design, Stefan Krastanov, Chao Shen and
Victor Albert for discussions on universal control and
Steve Flammia and Robin Blume-Kohout for advice about tomography.
This research was supported by the U.S. Army Research Office (W911NF-14-1-011).
P.R. was supported by the U.S. Air Force Office of Scientific Research (FA9550-15-1-0015),
L.J. by the Alfred P. Sloan Foundation and the Packard Foundation.
Facilities use was supported by the Yale Institute for Nanoscience and Quantum Engineering (YINQE),
the Yale SEAS cleanroom, and the National Science Foundation (MRSECDMR-1119826).


\bibliographystyle{unsrt}
\bibliography{oc_paper}

\begin{thebibliography}{10}

\bibitem{Terhal:2015ks}
Barbara~M Terhal.
\newblock {Quantum error correction for quantum memories}.
\newblock {\em Rev. Mod. Phys.}, 87(2):307--346, April 2015.

\bibitem{Zhang:2012hl}
Jingfu Zhang, Raymond Laflamme, and Dieter Suter.
\newblock {Experimental Implementation of Encoded Logical Qubit Operations in a
  Perfect Quantum Error Correcting Code}.
\newblock {\em Phys. Rev. Lett.}, 109(10):100503, September 2012.

\bibitem{Nigg:2014eb}
D~Nigg, M~M{\"u}ller, E~A Martinez, P~Schindler, M~Hennrich, T~Monz, M~A
  Martin-Delgado, and R~Blatt.
\newblock {Quantum computations on a topologically encoded qubit}.
\newblock {\em Science}, 345(6194):302--305, July 2014.

\bibitem{Mirrahimi:2014js}
Mazyar Mirrahimi, Zaki Leghtas, Victor~V Albert, Steven Touzard, Robert~J
  Schoelkopf, Liang Jiang, and Michel~H Devoret.
\newblock {Dynamically protected cat-qubits: a new paradigm for universal
  quantum computation}.
\newblock {\em New J. Phys.}, 16(4), 2014.

\bibitem{ofek2016demonstrating}
Nissim Ofek, Andrei Petrenko, Reinier Heeres, Philip Reinhold, Zaki Leghtas,
  Brian Vlastakis, Yehan Liu, Luigi Frunzio, SM~Girvin, Liang Jiang,
  M~Mirrahimi, M~H Devoret, and R~J Schoelkopf.
\newblock Extending the lifetime of a quantum bit with error correction in
  superconducting circuits.
\newblock {\em Nature}, advance online publication, Jul 2016.

\bibitem{Khaneja:2005jm}
Navin Khaneja, Timo Reiss, Cindie Kehlet, Thomas Schulte-Herbr{\"u}ggen, and
  Steffen~J Glaser.
\newblock {Optimal control of coupled spin dynamics: design of NMR pulse
  sequences by gradient ascent algorithms}.
\newblock {\em J. Mag. Res.}, 172(2):296--305, February 2005.

\bibitem{chiaverini2004realization}
J~Chiaverini, D~Leibfried, T~Schaetz, MD~Barrett, RB~Blakestad, J~Britton,
  WM~Itano, JD~Jost, E~Knill, C~Langer, et~al.
\newblock Realization of quantum error correction.
\newblock {\em Nature}, 432(7017):602--605, 2004.

\bibitem{Cramer:2016gv}
J~Cramer, N~Kalb, M~A Rol, B~Hensen, M~S Blok, M~Markham, D~J Twitchen,
  R~Hanson, and T~H Taminiau.
\newblock {Repeated quantum error correction on a continuously encoded qubit by
  real-time feedback}.
\newblock {\em Nature Communications}, 7:11526, May 2016.

\bibitem{Kelly:2015gi}
J~Kelly, R~Barends, Austin~G Fowler, A~Megrant, E~Jeffrey, T~C White, D~Sank,
  J~Y Mutus, B~Campbell, Yu~Chen, Z~Chen, B~Chiaro, A~Dunsworth, I~C Hoi,
  C~Neill, P~J~J O'Malley, C~Quintana, P~Roushan, A~Vainsencher, J~Wenner,
  Andrew~N Cleland, and John~M Martinis.
\newblock {State preservation by repetitive error detection in a
  superconducting quantum circuit}.
\newblock {\em Nature}, 519(7541):66--69, March 2015.

\bibitem{riste2015detecting}
D~Rist{\`e}, S~Poletto, M~Z Huang, A~Bruno, V~Vesterinen, O~P Saira, and
  L~DiCarlo.
\newblock {Detecting bit-flip errors in a logical qubit using stabilizer
  measurements}.
\newblock {\em Nature Communications}, 6:6983, April 2015.

\bibitem{corcoles2015demonstration}
A~D C{\'o}rcoles, Easwar Magesan, Srikanth~J Srinivasan, Andrew~W Cross,
  M~Steffen, Jay~M Gambetta, and Jerry~M Chow.
\newblock {Demonstration of a quantum error detection code using a square
  lattice of four superconducting qubits}.
\newblock {\em Nature Communications}, 6:6979, April 2015.

\bibitem{Gottesman:2001jb}
Daniel Gottesman, Alexei Kitaev, and John Preskill.
\newblock {Encoding a qubit in an oscillator}.
\newblock {\em Phys. Rev. A}, 64(1):012310, June 2001.

\bibitem{Michael:2016uu}
Marios~H Michael, Matti Silveri, R~T Brierley, Victor~V Albert, Juha
  Salmilehto, Liang Jiang, and Steven~M Girvin.
\newblock {New Class of Quantum Error-Correcting Codes for a Bosonic Mode}.
\newblock {\em Phys. Rev. X}, 6(3):031006, July 2016.

\bibitem{Law:1996cj}
C~K Law and J~H Eberly.
\newblock {Arbitrary Control of a Quantum Electromagnetic Field}.
\newblock {\em Phys. Rev. Lett.}, 76(7):1055--1058, February 1996.

\bibitem{Hofheinz:2009ba}
Max Hofheinz, H~Wang, M~Ansmann, Radoslaw~C Bialczak, Erik Lucero, M~Neeley,
  A~D O{\textquoteright}Connell, D~Sank, J~Wenner, John~M Martinis, and
  Andrew~N Cleland.
\newblock {Synthesizing arbitrary quantum states in a superconducting
  resonator}.
\newblock {\em Nature}, 459(7246):546--549, 2009.

\bibitem{Krastanov:2015gn}
Stefan Krastanov, Victor~V Albert, Chao Shen, Chang-Ling Zou, Reinier~W Heeres,
  Brian Vlastakis, Robert~J Schoelkopf, and Liang Jiang.
\newblock {Universal control of an oscillator with dispersive coupling to a
  qubit}.
\newblock {\em Phys. Rev. A}, 92(4):040303, October 2015.

\bibitem{Nigg:2014dl}
Simon~E Nigg.
\newblock {Deterministic Hadamard gate for microwave cat-state qubits in
  circuit QED}.
\newblock {\em Phys. Rev. A}, 89(2):022340, February 2014.

\bibitem{Vlastakis:2013ju}
Brian Vlastakis, Gerhard Kirchmair, Zaki Leghtas, Simon~E Nigg, Luigi Frunzio,
  Steven~M Girvin, Mazyar Mirrahimi, Michel~H Devoret, and Robert~J Schoelkopf.
\newblock {Deterministically Encoding Quantum Information Using 100-Photon
  Schr{\"o}dinger Cat States}.
\newblock {\em Science}, 342(6158):607--610, November 2013.

\bibitem{Heeres:2015kr}
Reinier~W Heeres, Brian Vlastakis, Eric Holland, Stefan Krastanov, Victor~V
  Albert, Luigi Frunzio, Liang Jiang, and Robert~J Schoelkopf.
\newblock {Cavity State Manipulation Using Photon-Number Selective Phase
  Gates}.
\newblock {\em Phys. Rev. Lett.}, 115(13):137002, September 2015.

\bibitem{deFouquieres:2011wm}
P~de~Fouquieres, SG~Schirmer, SJ~Glaser, and Ilya Kuprov.
\newblock Second order gradient ascent pulse engineering.
\newblock {\em J. Mag. Res.}, 212(2):412--417, 2011.

\bibitem{dolde2014high}
Florian Dolde, Ville Bergholm, Ya~Wang, Ingmar Jakobi, Boris Naydenov,
  S{\'e}bastien Pezzagna, Jan Meijer, Fedor Jelezko, Philipp Neumann, Thomas
  Schulte-Herbr{\"u}ggen, Jacob Biamonte, and J{\"o}rg Wrachtrup.
\newblock {High-fidelity spin entanglement using optimal control}.
\newblock {\em Nature Communications}, 5, February 2014.

\bibitem{Anderson:2015fu}
B~E Anderson, H~Sosa-Martinez, C~A Riofr{\'\i}o, Ivan~H Deutsch, and Poul~S
  Jessen.
\newblock {Accurate and Robust Unitary Transformations of a High-Dimensional
  Quantum System}.
\newblock {\em Phys. Rev. Lett.}, 114(24):240401, June 2015.

\bibitem{MattCav}
Matthew Reagor, Wolfgang Pfaff, Christopher Axline, Reinier~W. Heeres, Nissim
  Ofek, Katrina Sliwa, Eric Holland, Chen Wang, Jacob Blumoff, Kevin Chou,
  Michael~J. Hatridge, Luigi Frunzio, Michel~H. Devoret, Liang Jiang, and
  Robert~J. Schoelkopf.
\newblock Quantum memory with millisecond coherence in circuit qed.
\newblock {\em Phys. Rev. B}, 94:014506, Jul 2016.

\bibitem{sun2014tracking}
L~Sun, A~Petrenko, Z~Leghtas, B~Vlastakis, G~Kirchmair, KM~Sliwa, A~Narla,
  M~Hatridge, S~Shankar, J~Blumoff, L~Frunzio, M~Mirrahimi, Devoret~M H, and
  R~J Schoelkopf.
\newblock Tracking photon jumps with repeated quantum non-demolition parity
  measurements.
\newblock {\em Nature}, 511:444--448, 2014.

\bibitem{magesan2011scalable}
Easwar Magesan, Jay~M Gambetta, and Joseph Emerson.
\newblock {Scalable and Robust Randomized Benchmarking of Quantum Processes}.
\newblock {\em Phys. Rev. Lett.}, 106(18):180504, May 2011.

\bibitem{Magesan:2012gu}
Easwar Magesan, Jay~M Gambetta, Blake~R Johnson, Colm~A Ryan, Jerry~M Chow,
  Seth~T Merkel, Marcus~P da~Silva, George~A Keefe, Mary~B Rothwell, Thomas~A
  Ohki, Mark~B Ketchen, and M~Steffen.
\newblock {Efficient Measurement of Quantum Gate Error by Interleaved
  Randomized Benchmarking}.
\newblock {\em Phys. Rev. Lett.}, 109(8):080505, August 2012.

\bibitem{byrd1995limited}
Richard~H Byrd, Peihuang Lu, Jorge Nocedal, and Ciyou Zhu.
\newblock A limited memory algorithm for bound constrained optimization.
\newblock {\em SIAM Journal on Scientific Computing}, 16(5):1190--1208, 1995.

\bibitem{Najfeld:1995fo}
I~Najfeld and T~F Havel.
\newblock {Derivatives of the Matrix Exponential and Their Computation}.
\newblock {\em Advances in Applied Mathematics}, 16(3):321--375, September
  1995.

\bibitem{Motzoi:2011hb}
F~Motzoi, Jay~M Gambetta, S~T Merkel, and F~K Wilhelm.
\newblock {Optimal control methods for rapidly time-varying Hamiltonians}.
\newblock {\em Phys. Rev. A}, 84(2):022307, August 2011.

\bibitem{Egger:2014di}
D~J Egger and F~K Wilhelm.
\newblock {Adaptive Hybrid Optimal Quantum Control for Imprecisely
  Characterized Systems}.
\newblock {\em Phys. Rev. Lett.}, 112(24):240503, June 2014.

\bibitem{Kelly:2014fh}
J~Kelly, R~Barends, B~Campbell, Y~Chen, Z~Chen, B~Chiaro, A~Dunsworth, Austin~G
  Fowler, I~C Hoi, E~Jeffrey, A~Megrant, J~Mutus, C~Neill, P~J~J O'Malley,
  C~Quintana, P~Roushan, D~Sank, A~Vainsencher, J~Wenner, T~C White, Andrew~N
  Cleland, and John~M Martinis.
\newblock {Optimal Quantum Control Using Randomized Benchmarking}.
\newblock {\em Phys. Rev. Lett.}, 112(24):240504, June 2014.

\bibitem{kirchmair2013observation}
Gerhard Kirchmair, Brian Vlastakis, Zaki Leghtas, Simon~E Nigg, Hanhee Paik,
  Eran Ginossar, Mazyar Mirrahimi, Luigi Frunzio, Steven~M Girvin, and Robert~J
  Schoelkopf.
\newblock Observation of quantum state collapse and revival due to the
  single-photon kerr effect.
\newblock {\em Nature}, 495(7440):205--209, 2013.

\end{thebibliography}

\clearpage
\onecolumngrid
\setcounter{figure}{0}
\setcounter{table}{0}
\setcounter{section}{0}
\begin{center}
\textbf{Supplementary Material}
\end{center}
\renewcommand\thefigure{S\arabic{figure}}
\renewcommand\thetable{S\Roman{table}}

\section{System Hamiltonian}
Here we give the full system Hamiltonian to the precision with which we have characterized it.
We denote the annihilation operator corresponding to the oscillator (transmon) mode with
$\hat{a}$ ($\hat{b}$). Breaking down the system Hamiltonian into components representing
the individual modes, their interactions, as well as driving terms, we can write

\begin{align}
H(t) &= H_\text{oscillator} + H_\text{transmon} +  H_\text{interaction} + H_\text{drive}(t)\\
H_\text{oscillator}/\hbar &=  \omega_C \hat{a}^\dagger \hat{a} + \frac{K}{2}(\hat{a}^\dagger)^2 \hat{a}^2\\
H_\text{transmon}/\hbar &= \omega_T \hat{b}^\dagger \hat{b} +
					 \frac{\alpha}{2} (\hat{b}^\dagger)^2 \hat{b}^2\\
H_\text{interaction}/\hbar &= \chi \hat{a}^\dagger \hat{a} \hat{b}^\dagger \hat{b} +
						\frac{\chi'}{2}\hat{b}^\dagger \hat{b} (\hat{a}^\dagger)^2 \hat{a}^2\\
H_\text{drive}(t)/\hbar &=  \epsilon_C(t)\hat{a} + \epsilon_T(t)\hat{b} +  \text{h.c.}
\end{align}

When simulating how known decoherence sources should impact the fidelity of our operations,
we use a Markovian Lindblad master equation of the form:

\begin{align}
\frac{\partial}{\partial t}\rho(t) &= \frac{-i}{\hbar}[H(t), \rho(t)] +
\left(
\frac{1}{T_{1,C}}D[\hat{a}] +
\frac{1}{T_{1,T}}D[\hat{b}] +
\frac{1}{T_\phi}D[\hat{b}^\dagger \hat{b}]
\right)(\rho(t))\\
D[a](\rho) &= a\rho a^\dagger - \frac{1}{2}\{a^\dagger a, \rho\}
\end{align}

The measured values for each of these system parameters are shown in
table~\ref{table:sys_params}.

\begin{table}
\begin{tabular}{l l l}
System Parameter & Hamiltonian Term & Parameter Value \\
\hline
Transmon frequency & $\omega_T \hat{b}^\dagger \hat{b}$ & $2\pi \times 5664.0\unit{MHz}$\\
Oscillator frequency & $\omega_C \hat{a}^\dagger \hat{a} $ & $2\pi \times 4452.6 \unit{MHz}$\\
Dispersive shift & $\chi \hat{a}^\dagger \hat{a} \hat{b}^\dagger \hat{b}$ & $2\pi \times -2194 \pm 3 \unit{kHz}$\\
Transmon anharmonicity & $\frac{\alpha}{2}(\hat{b}^\dagger)^2 \hat{b}^2$ & $2\pi \times -236 \unit{MHz}$\\
Oscillator anharmonicity (Kerr)& $\frac{K}{2}(\hat{a}^\dagger)^2 \hat{a}^2$ & $2\pi \times-3.7 \pm 0.1 \unit{kHz}$\\
Second order dispersive shift & $\frac{\chi'}{2}(\hat{a}^\dagger)^2 \hat{a}^2 \hat{b}^\dagger \hat{b}$ & $2\pi\times -19.0 \pm 0.4 \unit{kHz}$\\
Transmon relaxation & $\frac{1}{T_1} D[\hat{b}]$ & $170 \pm 10 \mu s$\\
Transmon dephasing & $\frac{1}{T_\phi} D[\hat{b}^\dagger \hat{b}]$ & $43 \pm 5 \mu s$\\
Oscillator relaxation & $\frac{1}{T_\text{cav}} D[\hat{a}]$ & $2.7 \pm 0.1 ms$\\
\end{tabular}
\caption{\label{table:sys_params}\textbf{Measured system parameters}
The dispersive shift and its second order correction are determined using
transmon spectroscopy experiments (Fig.~\ref{chi_spec}). The oscillator anharmonicity is determined by fitting
a set of Wigner functions after different lengths of free evolution time.
}
\end{table}

\begin{figure}
\centerline{\includegraphics{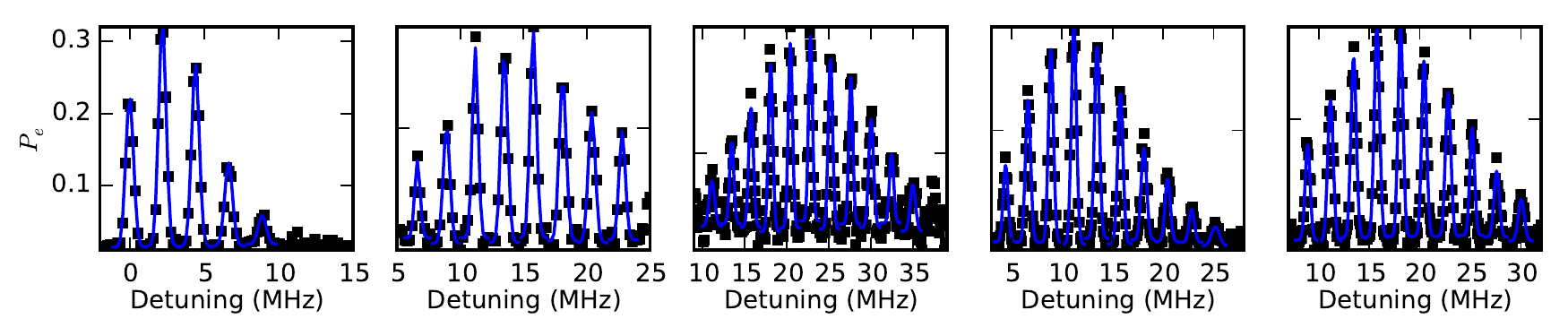}}
\centerline{\includegraphics{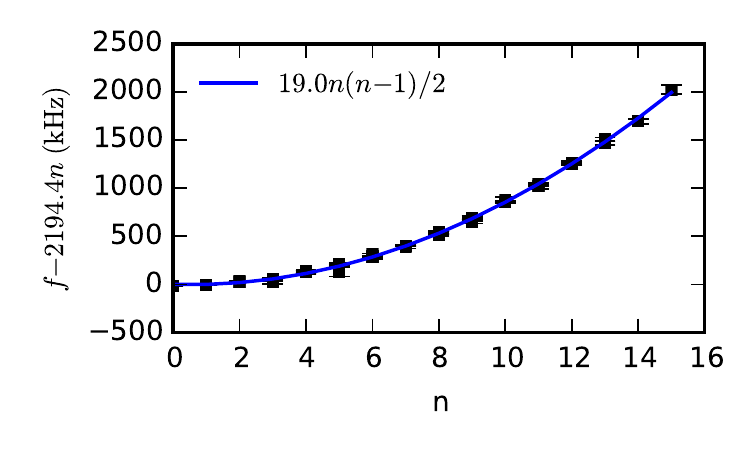}}
\caption{\label{chi_spec}
\textbf{Dispersive shift measurement.}
The dispersive shift $\chi$ and its second order correction
term $\chi'$ are determined from transmon spectroscopy experiments with
several different displacements (top).
Each peak is fit to a Gaussian and the resulting center frequencies
are fit using a quadratic model.
}

\end{figure}
\section{GRAPE implementation}
We define operations on our system in terms of a set of simultaneous state
transfers, i.e.\ the operation should, for each $i$, take
the initial state $\ket{\psi_\text{init}^{(i)}}$ to the corresponding final
state $\ket{\psi_\text{final}^{(i)}}$.
In order to prepare a desired operation on the joint oscillator-transmon Hilbert
space, we use \textsc{grape} to maximize the (coherent) average fidelity of these state
transfers over the controls
$\bm{\epsilon}(t) \equiv \left(\epsilon_C(t),\,\epsilon_T(t)\right)$:
\begin{equation}
\label{eq:state-transfer}
\underset{\bm{\epsilon}(t)}{\text{maximize}}\;\;
\mathcal{F}\left(\bm{\epsilon}(t)\right)
\end{equation}
\begin{equation}
\mathcal{F}\left(\bm{\epsilon}(t)\right) =
\left|
\sum_{i}\bra{\psi_\text{final}^{(i)}}
U\left(T,\bm{\epsilon}(t)\right)\ket{\psi_\text{init}^{(i)}}
\right|^2,
\end{equation}
where the unitary $U$ defined by the waveforms $\bm{\epsilon}(t)$
is given by the time-ordered exponential of the Hamiltonian up to some
final time $T$,
\begin{equation}
U(T,\bm{\epsilon}(t)) =
\mathcal{T}\exp\left(
-\int_0^T\!\!\!\!\mathrm{d}t\,
H\left(\bm{\epsilon}(t)\right)
\right).
\end{equation}
To make the problem numerically tractable, $\bm{\epsilon}(t)$ is
represented as a piecewise constant function with $N = T/\Delta t$ steps
of length $\Delta t = 2\unit{ns}$, corresponding to the time resolution of our
arbitrary waveform generator.
\begin{align}
U\left(T,\bm{\epsilon}(t)\right) &= U_N U_{N-1} \cdots U_2 U_1\\
U_k &= \exp\left(\frac{i\Delta t}{\hbar} H(\bm{\epsilon}(k \Delta t)\right)
\end{align}
Using $4$ parameters per time point (real and imaginary components
of the oscillator and transmon drives) and $N=550$ time points representing
the 1.1 $\mu$s pulse, there are 2200 parameters to optimize over.
In order to carry out a numerical optimization with such a large number
of parameters, it is crucial that one can efficiently calculate the
gradient of the optimized function with respect to all of its parameters.
In this case it is possible to use Quasi-Newton optimization algorithms,
such as L-BFGS \cite{byrd1995limited} in order to optimize the function
with many fewer function evaluations. We can simplify the calculation of the
gradient as follows:
\begin{align}
\frac{\partial \mathcal{F}}{\partial \epsilon_i(k\Delta t)} &=
2 \left(
\text{Re}(v) \text{Re}\left(\frac{\partial v}{\partial \epsilon_i(k\Delta t)}\right) +
\text{Im}(v) \text{Im}\left(\frac{\partial v}{\partial \epsilon_i(k\Delta t)}\right)
\right)\\
v &\equiv \sum_{i}\bra{\psi_\text{final}^{(i)}}
U\left(T,\bm{\epsilon}(t)\right)\ket{\psi_\text{init}^{(i)}}\\
\frac{\partial v}{\partial \epsilon_i(k\Delta t)} &=
\sum_{i}\bra{\psi_\text{final}^{(i)}}
\frac{\partial U\left(T,\bm{\epsilon}(t)\right)}
{\partial\epsilon_i(k\Delta t)}\ket{\psi_\text{init}^{(i)}}\\
&=
\sum_{i}\bra{\psi_\text{final}^{(i)}}
U_N\cdots U_{k+1}
\frac{\partial U_k}{\partial \epsilon_i(k\Delta t)}
U_{k-1}\cdots U_1
\ket{\psi_\text{init}^{(i)}}
\end{align}

Therefore, the calculation of the gradient can be reduced to computing the states
$U_{k-1}\cdots U_1\ket{\psi_\text{init}^{(i)}}$,
$U_{k+1}^\dagger\cdots U_N^\dagger\ket{\psi_\text{final}^{(i)}}$
as well as the gradient of the step propagator
$\frac{\partial U_k}{\partial \epsilon_i(k\Delta t)}$.
The states can be stored from the evaluation of the fidelity itself,
and there are several efficient ways of evaluating the gradient of the
propagator \cite{Najfeld:1995fo}.

The optimization problem defined by equation~\ref{eq:state-transfer} is
generally underdetermined, i.e. there are many solutions
$\bm{\epsilon}(t)$ which achieve equally high fidelities. Therefore,
we can add additional terms to the optimization cost function, such that the
resulting solution optimizes against several other desiderata. For a set of
constraints on the solution $\{g_i \geq 0\}$, where ideally
$g_i\left(\bm{\epsilon}(t)\right) = 0$, we can associate a Lagrange multiplier
$\lambda_i$, and modify our optimization to read:
\begin{equation}
\underset{\bm{\epsilon}(t)}{\text{maximize}}\;\;
\mathcal{F}\left(\bm{\epsilon}(t)\right) - \sum_i\lambda_i g_i\left(\bm{\epsilon}(t)\right)
\end{equation}
The values $\lambda_i$ are chosen by trial-and-error, set to be just large enough
that the violation of the constraint upon termination is within acceptable levels.
For instance, since the output power of our AWG is limited, the pulse must obey
$\epsilon(t) \leq \epsilon_\text{max}$ for all $t$. We can construct a penalty term
of the form
\begin{align}
g_\text{amplitude}\left(\epsilon(t)\right) &=
\int\!\!\mathrm{d}t\left(\left|\epsilon(t)\right| - \epsilon_\text{max}\right)^2
\Theta\left(\left|\epsilon(t)\right| - \epsilon_\text{max}\right)\\
&= \sum_n \left(\left|\epsilon(n\Delta t)\right| - \epsilon_\text{max}\right)^2
\Theta\left(\left|\epsilon(n\Delta t)\right| - \epsilon_\text{max}\right)
\end{align}

Since the transfer function of the lines between the AWG and the experimental system
becomes more and more uncertain as one moves further away from resonance, it is also
desirable to minimize the bandwidth of the applied pulses, we do this in two ways.
First, we create a penalty term of the form

\begin{align}
g_\text{derivative}\left(\epsilon(t)\right) &=
\int\!\!\mathrm{d}t\left(\frac{\partial \epsilon(t)}{\partial t}\right)^2 \label{deriv_pen_cont}\\
&\rightarrow \sum_n \left(\epsilon((n+1)\Delta t) - \epsilon(n\Delta t)\right)^2 \label{deriv_pen_pwc},
\end{align}
where equation~\ref{deriv_pen_pwc} is the appropriate equivalent of equation~\ref{deriv_pen_cont}
for a piecewise constant function. Additionally, we enforce a hard cutoff on the minimum and
maximum frequencies allowed in the solution by reparametrizing the optimization problem in
terms of the Fourier transform of the pulses \cite{Motzoi:2011hb}:
\begin{align}
\underset{\tilde{\bm{\epsilon}}(\omega)}{\text{maximize}}&\;\;
\mathcal{F}\left(\bm{\epsilon}(t)\right) - \sum_i\lambda_i g_i\left(\bm{\epsilon}(t)\right)\\
\text{such that}&\;\; \tilde{\bm{\epsilon}}(\omega) = 0\;\text{when}\;\omega < \omega_\text{min}\;
\text{or}\;\omega > \omega_\text{max}\nonumber
\end{align}

Since computer memory is finite, we are forced to choose a photon number truncation $N$
such that the operator $\hat{a}$ becomes a $N\times N$ matrix. When we do this, we are
in effect replacing our infinite-dimensional oscillator with a finite-dimensional qudit.
This replacement is only valid if all of the system dynamics relevant for the desired state transfers
occurs within the $\{\ket{0},\ldots,\ket{N-1}\}$ subspace. For generic applied drives this is not the
case. In order to enforce this property, we modify the optimization problem to find a solution
which operates identically under several different values of $N$. Writing the fidelity
as computed with a truncation $N$ as $\mathcal{F}_N$, we have:

\begin{align}
\underset{\tilde{\bm{\epsilon}}(\omega)}{\text{maximize}}\;\;
\left(\sum_k\mathcal{F}_{N+k}\left(\bm{\epsilon}(t)\right)\right) -
\left(\sum_i\lambda_i g_i\left(\bm{\epsilon}(t)\right)\right)
\end{align}
To enforce that the behavior is identical in the different truncations, we add the penalty term
\begin{equation}
g_\text{discrepancy}\left(\bm{\epsilon(t)}\right) =
\sum_{k_1 \neq k_2} \left(
\mathcal{F}_{N+k_1}\left(\bm{\epsilon}(t)\right) -
\mathcal{F}_{N+k_2}\left(\bm{\epsilon}(t)\right)\right)^2
\end{equation}

The choice of $N$ determines the maximum
photon number population which can be populated during the pulse, and figures in determining the minimum
time necessary for the operation (faster pulses can be achieved with higher $N$).

\begin{figure}
\centerline{\includegraphics[width=19cm]{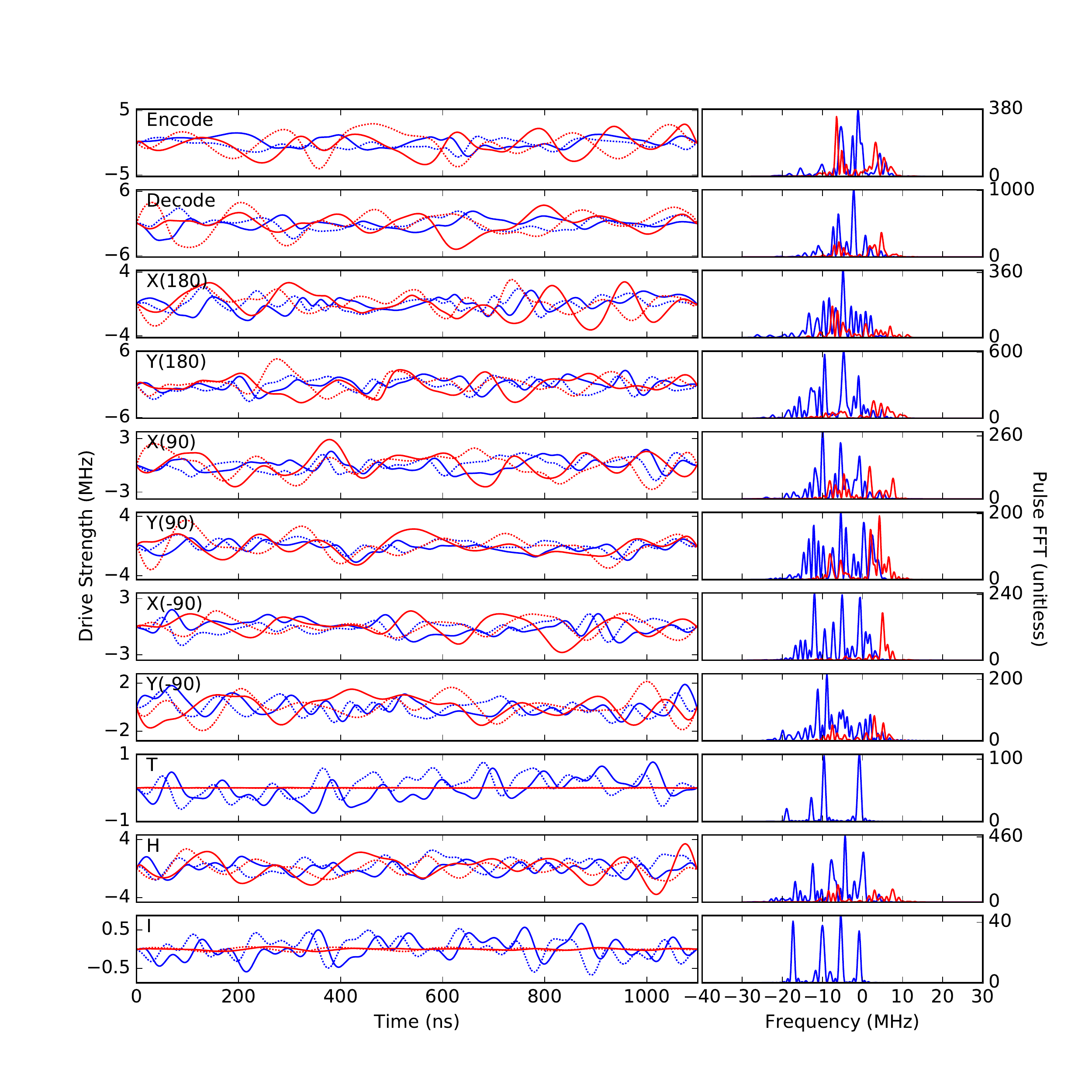}}
\caption{\label{pulse_plot}
\textbf{Optimized pulse waveforms.}
In the first column, we plot the complex waveforms
$\epsilon_T(t)$ and $\epsilon_C(t)$. In the second
column, we show the Fourier spectrum $|\tilde{\epsilon}(\omega)|^2$.
Blue (red) lines
correspond to drives on the transmon (oscillator).
Solid (dotted) lines correspond to the in-phase (quadrature) component of the drive.
Note that the I and the T gate do not have to change the photon number
distribution, but only have to apply different phases to each Fock state
component. This can be done by manipulating the transmon \cite{Heeres:2015kr} only;
\textsc{grape} finds a solution with a very small oscillator drive amplitude as well.
}
\end{figure}

\FloatBarrier
\section{Measurement setup}
\begin{figure*}[tbph]
\includegraphics{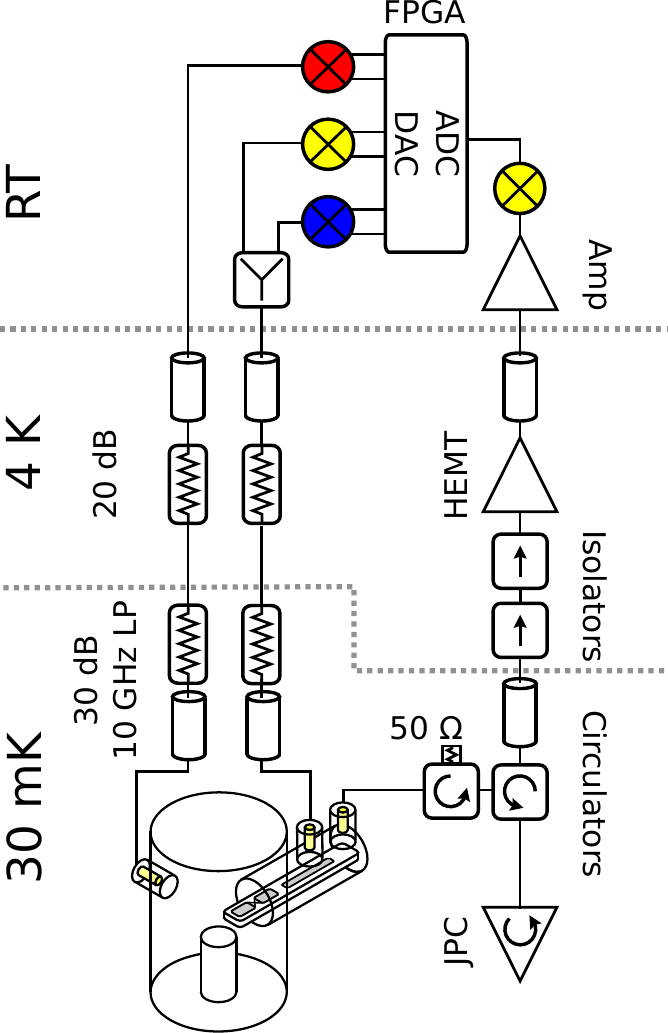}
\caption{\label{msmt_setup}\textbf{Measurement setup.}
An FPGA controller (2x Innovative Integration X6-1000M in VPXI-ePC chassis)
generates 3 pairs of I/Q waveforms using 500 Msample/s digital to analog converters (DAC).
Each pair is upconverted using an I/Q mixer
(Marki IQ-0307-LXP or IQ-0618-LXP depending on the frequency).
The color of the mixer indicates the local oscillator: red for the storage,
yellow for the readout and blue for the transmon. To prevent problems
due to mixer leakage, each local oscillator is set $50 \unit{MHz}$ above
the desired frequency and single-sideband modulation is used.
Proper attenuation at each temperature stage is crucial to
thermalize the black-body radiation
from the $50 \Omega$ environment. Additional low-pass filters
(K\&L250-10000 and home-built eccosorb) protect the sample from
spurious high-frequency components. The output chain consists of
a Josephson Parametric Converter (JPC), which reflects the input
signal with $\sim 20 \unit{dB}$ of gain (bandwidth $\sim 6 \unit{MHz}$).
The circulators (Pamtech XTE0812KC) prevent the amplified signal from going back
to the sample and direct it through 2 isolators (Pamtech CWJ0312KI)
to a HEMT-amplifier (Low Noise Factory LNF-LNR1\_12A).
Finally, an image reject mixer (Marki SSB-0618) converts the RF signal
back to the intermediate frequency ($50 \unit{MHz}$). The FPGA
samples the signal using a 1 Gsample/s analog to digital converter (ADC),
demodulates and integrates to give one bit of information indicating
whether the transmon was in $\ket{g}$ or $\ket{e}$.}
\end{figure*}
\FloatBarrier
\section{System preparation}
\begin{figure*}[tbph]
\centerline{\includegraphics[width=17cm]{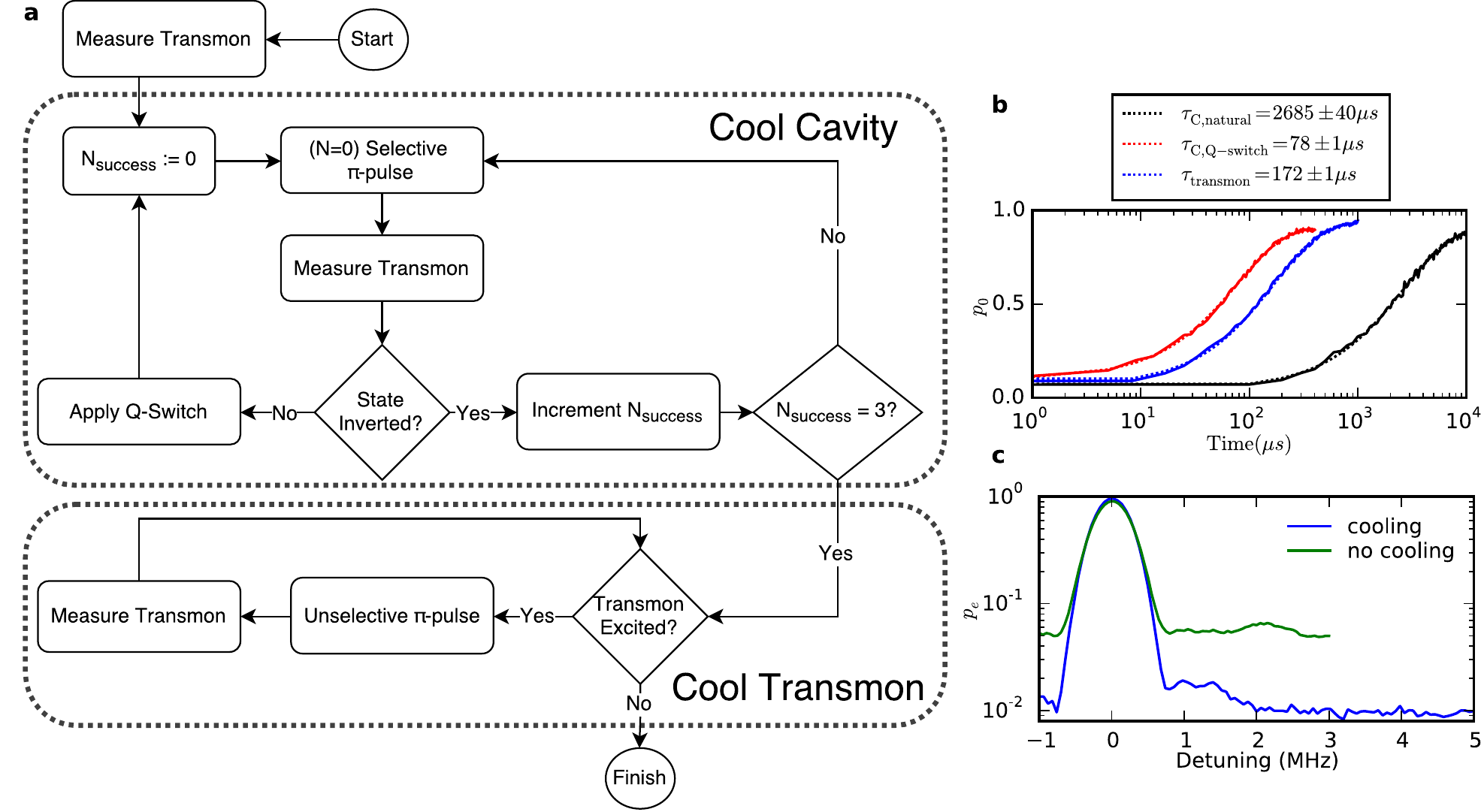}}
\caption{\label{system_prep}\textbf{System preparation.}
\textbf{a,} System preparation protocol to cool the oscillator as well as the transmon.
\textbf{b,} Lifetime of the transmon and oscillator. For the
oscillator we prepare the Fock state $\ket{1}$ using an optimal control pulse
and show a natural decay curve as well as one with Q-switching pumps applied.
\textbf{c,}
  Transmon spectroscopy data after system preparation.
  The ''cooling'' (''no-cooling'') curve are with (without) the feedback-cooling protocol.
  Photons in the storage (readout) oscillator show up as a peak around
  $\chi_s \approx 2 \unit{MHz}$ ($\chi_r \approx 1 \unit{MHz}$).
}
\end{figure*}
The system is initialized by cooling of both the storage resonator (typical steady-state population $\sim 2\%$)
and the transmon (steady-state population $\sim 5 \%$) using measurement-based feedback.
The protocol is detailed in figure~\ref{system_prep}. It proceeds by
first establishing that the oscillator is in its ground state, and finishes by
ensuring that the transmon is in its ground state (figure~\ref{system_prep}a).
If it is determined that the oscillator is not empty, a set of "Q-Switching" drives
is applied which effectively couples the storage mode to the short-lived readout mode
(figure~\ref{system_prep}b). The drives consist of strong tones applied at
$\omega_{C} + \Delta$ and $\omega_\text{RO} + \Delta$ with $\Delta = 40\unit{MHz}$.
The effectiveness of this strategy can be seen from the transmon spectroscopy traces
(figure~\ref{system_prep}c).
The transmon population is reduced to $\sim 1\%$ and the storage resonator population
is $\ll 1\%$. A residual population of the readout resonator of about $1 \%$ is
visible as a peak around $1\unit{MHz}$ detuning. Additionally, this cooling protocol
allows for a dramatically increased experimental repetition rate, decreasing the
inter-experimental delay $\tau$ from $\tau \approx 18\unit{ms}$ to $\tau < 1\unit{ms}$.

\section{Empirical tuning}
\begin{figure*}[tbph]
  \mbox{\includegraphics{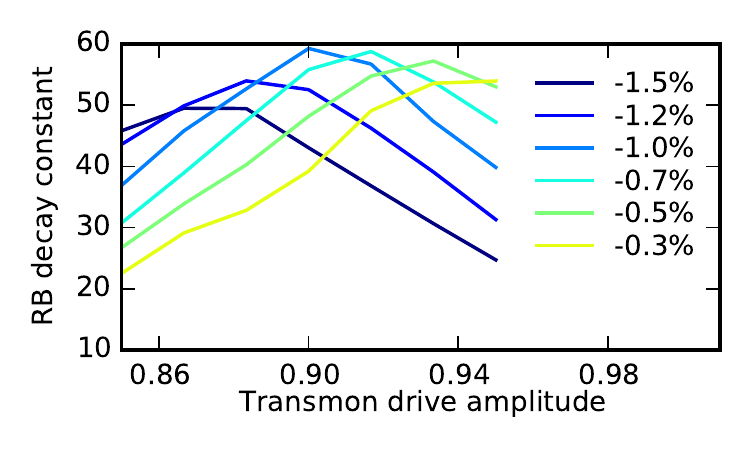}}
  \mbox{\includegraphics{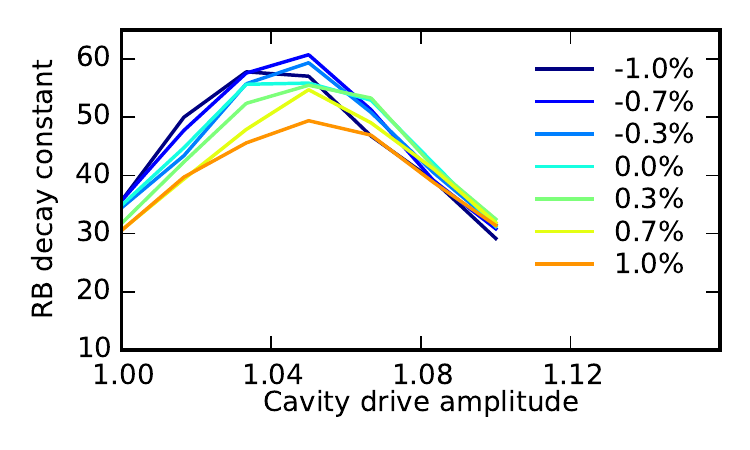}}
  \caption{\label{dispersion_tuneup}\textbf{Dispersion and amplitude optimization}
The randomized benchmarking decay constant versus transmon drive amplitude for
several different dispersion values (in \% per MHz). Because of the spectral content
of the pulses, the amplitude might have to be corrected when the dispersion value is adjusted.}
\end{figure*}
\begin{figure*}[tbph]
  \mbox{\includegraphics{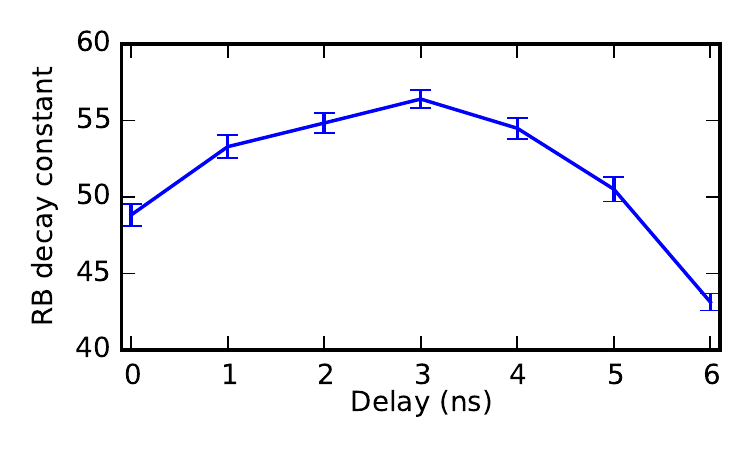}}
  \caption{\label{delay_tuneup}\textbf{Delay optimization} }
\end{figure*}

We use the randomized benchmarking protocol to perform fine tuning of
the resulting pulse waveforms \cite{Egger:2014di,Kelly:2014fh}.
Since the cables and frequency
modulation setup between our waveform generator and our device are not spectrally
flat, we attempt to find a correction to the pulse by applying a linear amplitude
weighting in the frequency domain, i.e.\ Fourier transforming the waves to find
$\tilde{\epsilon}(\omega)$, transforming using the weighting coefficient $b$
and delay parameter $\tau$.
$\tilde{\epsilon}(\omega)\rightarrow (1 + b\omega e^{i\omega\tau})\tilde{\epsilon}(\omega)$,
and inverse Fourier transforming to find the corrected waves in the time domain.
We can empirically optimize the value of $b$ (figure~\ref{dispersion_tuneup})
and $\tau$ (figure~\ref{delay_tuneup}) using randomized benchmarking.

\section{Additional data}

\begin{figure*}[tbph]
\includegraphics{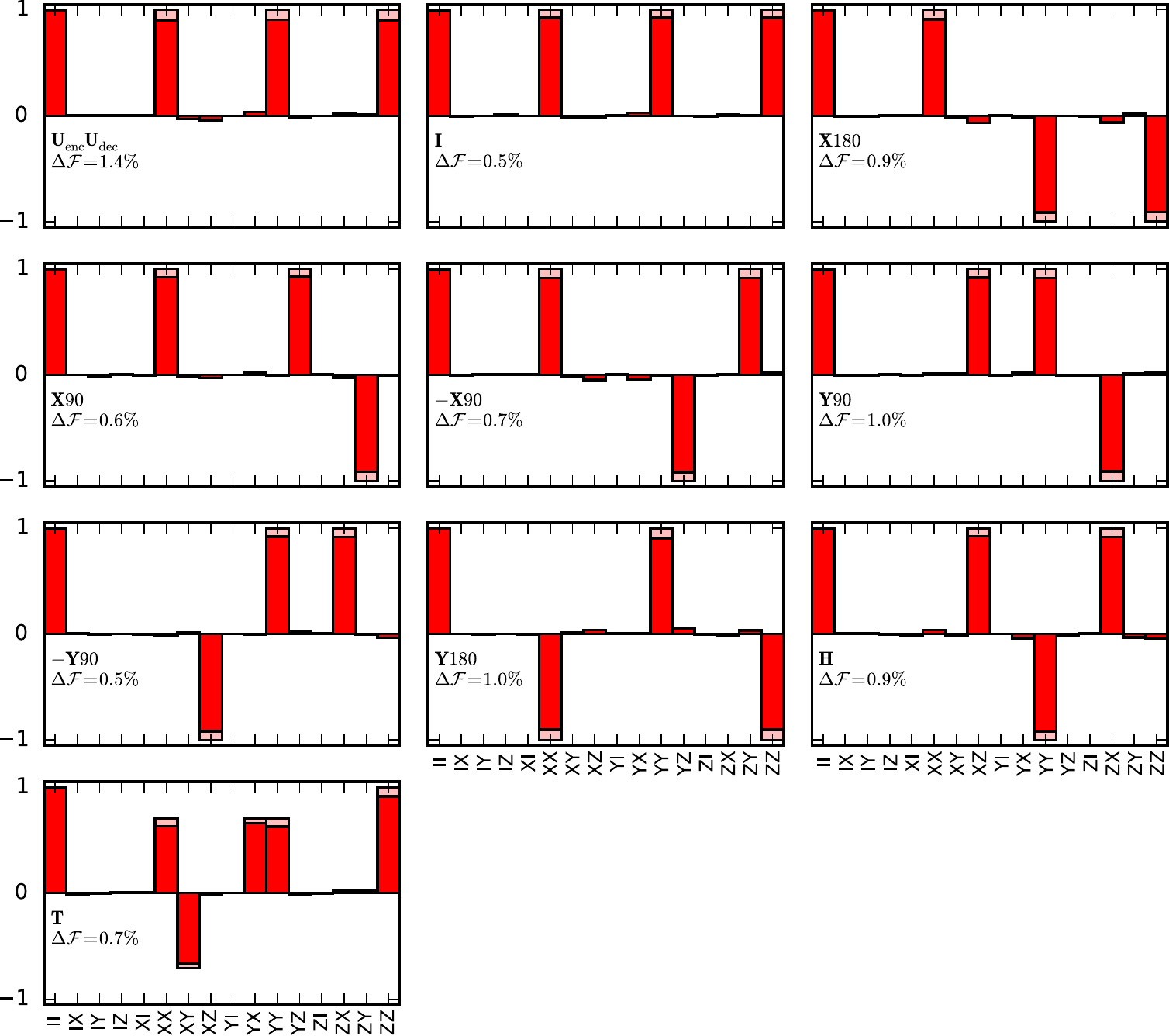}
\caption{\label{full_pt_data}\textbf{Full Process Tomography Results.}}
\end{figure*}

\begin{figure*}[tbph]
  \mbox{\includegraphics{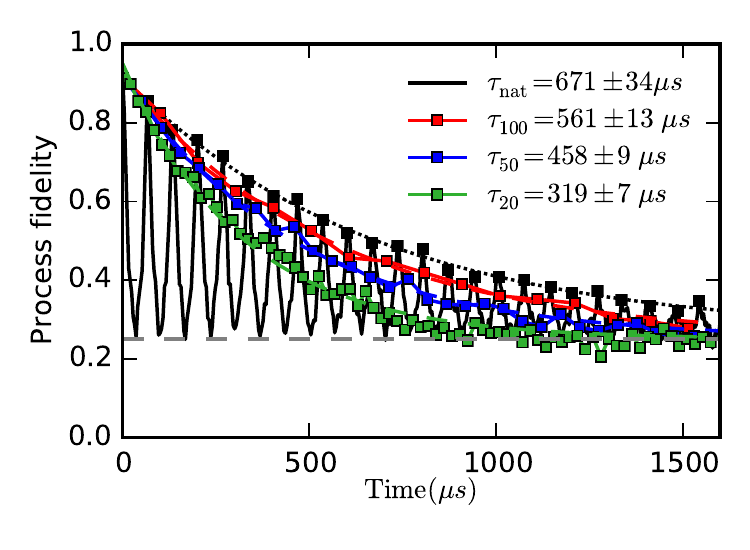}}
  \caption{\label{fid_vs_delay} \textbf{Lifetime of the cat-qubit}.
The long-lived resonator is not perfectly harmonic; its Kerr is $-3.7 \unit{kHz}$.
This nonlinearity will scramble the basis-states of our encoding under free evolution.
The black curve shows the process fidelity versus waiting time:
periodic revivals \cite{kirchmair2013observation} associated with rephasing of the
basis-states are clearly visible. Note that the revival periodicity is not $1/K$ due
to an intentional frequency detuning.
Our control pulses take this effect into account during their operating time.
Additionally, we can design a control pulse that corrects for the evolution associated
with some time $\Delta t$ of free evolution. The different curves in this plot are the result
of a stroboscopic Kerr-correction experiment for 20, 50 and 100 $\unit{\mu s}$ of free-evolution.
We can infer a Kerr-correction gate error of $\approx 1.7 \%$ per gate by extracting the
additional decay rate compared to the natural decay.}
\end{figure*}

\end{document}